\documentclass[conference]{IEEEtran}

\usepackage{cite}      

\usepackage{amsmath}   
\usepackage{amssymb}

\newcommand{\A}{\ensuremath{\mathcal A}}
\newcommand{\Ah}{\ensuremath{A_\hbar}}
\newcommand{\B}{\ensuremath{\mathcal B}}
\newcommand{\Bh}{\ensuremath{B_\hbar}}

\newcommand{\C}{\ensuremath{\mathbb C}}
\newcommand{\CC}{\ensuremath{\mathcal C}}

\newcommand{\E}{\ensuremath{\mathcal E}}

\newcommand{\h}{\hbar}
\renewcommand{\H}{\ensuremath{\mathcal H}}
\newcommand{\I}{\ensuremath{\mathcal I}}
\newcommand{\K}{\ensuremath{\mathcal K}}

\newcommand{\N}{\ensuremath{\mathbb N}}
\renewcommand{\O}{\ensuremath{\hat{\mathcal O}}}
\renewcommand{\P}{\ensuremath{\mathcal P}}

\newcommand{\R}{\ensuremath{\mathbb R}}

\newcommand\cospec{\operatorname{cospec}}

\newcommand{\cs}{{C^{*}}}

\renewcommand{\epsilon}{{\varepsilon}}

\newcommand{\ideal}{\vartriangleleft}
\newcommand{\id}{\operatorname{id}}

\newcommand\Interior{\operatorname{Interior}}

\newcommand{\Prob}{{\operatorname{Prob}}}

\newcommand{\range}{{\operatorname{range}}}

\newcommand\spec{\operatorname{spec}}

\newcommand{\supp}{{\operatorname{supp}}}
\newcommand\tensor{\otimes}
\newcommand{\tr}{{\operatorname{trace}}}

\newcommand\<{\langle}
\renewcommand\>{\rangle}

\newcommand\lcl{[\![}
\newcommand\rcl{]\!]}
\renewcommand\({(\!(}
\renewcommand\){)\!)}

\newtheorem{thm}{Theorem}[section] 
\newtheorem{dfn}[thm]{Definition}

\newtheorem{cor}[thm]{Corollary}

\newtheorem{ex}[thm]{Example}
\newtheorem{lem}[thm]{Lemma}

\newtheorem{prop}[thm]{Proposition}

\newtheorem{rmk}[thm]{Remark}
\newtheorem{thm*}{Theorem}

  {\addvspace{\bigskipamount}\noindent{\bf Example\ }}%
  {}

\newenvironment{tightlist}%
{\begin{list}{$\bullet$}{\parsep=0pt \itemsep=0pt 
                         \topsep=6pt \partopsep=\parskip}}%
{\end{list}}

\newcounter{ictr}

\newcounter{nctr}

\begin{document}
\title{Asymptotic Spectral Measures: Between Quantum Theory and E-theory}
\author{\authorblockN{Jody Trout}
\authorblockA{Department of Mathematics\\
Dartmouth College\\
Hanover, NH 03755\\
Email: jody.trout@dartmouth.edu}}


\maketitle

\begin{abstract}
We review the relationship between positive operator-valued
 measures (POVMs) in quantum measurement theory and asymptotic morphisms in the
$\cs$-algebra $E$-theory of Connes and Higson. The theory of
asymptotic spectral measures, as introduced by Martinez and Trout \cite{MT02}, is integrally related to positive asymptotic morphisms on locally compact spaces via an asymptotic Riesz Representation Theorem. Examples and applications
to quantum physics, including quantum noise models, semiclassical limits, pure spin one-half systems and quantum information processing will also be discussed.
\end{abstract}


%

\section{Introduction}

In the von Neumann Hilbert space model \cite{VN32} of quantum mechanics,
quantum observables are modeled as self-adjoint operators on the Hilbert 
space of states of the quantum system. The Spectral Theorem relates this
theoretical view of a quantum observable to the more 
operational one of a projection-valued measure (PVM or spectral measure) which
determines the probability distribution of the experimentally measurable values
of the observable.  In order to solve
foundational problems with the concept of measurement and to better analyze
unsharp results in experiments, this view  was
generalized to include positive operator-valued measures (POVMs).  Since the initial work of Jauch and
Piron \cite{JP67}, 
POV-measures have played an ever increasing role in both the foundations and
operational aspects of quantum physics  \cite{BGL95,Sch96}.

In this paper, we review the relationship between POVMs and asymptotic morphisms
in the operator algebra $E$-theory of 
Connes and Higson \cite{CH90}, which has already found many applications in
mathematics  \cite{Bla98,GHT00,Hig93,Trou99}, most notably to
classification problems in operator $K$-theory, index theory, representation
theory, geometry, and topology.  The basic ingredients of $E$-theory
are asymptotic morphisms, which are given by continuous families of functions
\begin{equation*}\{Q_\hbar\}_{\hbar \in (0,1]} : \A \to \B
\end{equation*}
 from a
$C^*$-algebra $\A$ to a $C^*$-algebra $\B$ that satisfy the axioms of a
$*$-homomorphism in the limit as the parameter $\hbar \to 0$. 
Asymptotic multiplicativity is a modern version of the Bohr-von Neumann
correspondence principle \cite{Land98} from quantization theory: For all $f, g$ in
$\A,$
\begin{equation*}
Q_\hbar(fg) - Q_\hbar(f)Q_\hbar(g) \to 0 \text{ as } \hbar \to 0.
\end{equation*}
It is then no surprise that quantization schemes may naturally define
asymptotic  morphisms, say, from the $C^*$-algebra $\A$ of classical observables
to the $C^*$-algebra $\B$ of quantum observables. 
See the papers \cite{Nag96,Nag97,Ros96} and the books \cite{Con94,GVF01} for more on the
connections between operator algebra $K$-theory, $E$-theory, and quantization.  

In \cite{MT02}, Martinez and Trout showed that there is a fundamental quantum-$E$-theory relationship by
introducing the concept of an {\it asymptotic spectral measure} (ASM or
asymptotic PVM)
$$\{A_\hbar\}_{\hbar \in (0,1]} : \Sigma \to \B(\H)$$
associated to a  measurable space $(X, \Sigma)$. (See Definition \ref{def:asm}.) Roughly,
this is a continuous
family of POV-measures which are ``asymptotically'' projective (or
quasiprojective)
as $\hbar \to 0$:
$$ \Ah(\Delta \cap \Delta') - \Ah(\Delta) \Ah(\Delta') \to 0 \text{ as  } \hbar \to 0$$
for certain measurable sets $\Delta, \Delta' \in \Sigma$.

Let $X$  be the locally compact phase space of a classical system. Let $C_0(X)$ denote the $C^*$-algebra
of continuous functions vanishing at infinity on $X$.  Let $\H$ be the Hilbert state space of the quantum version of the system. Let $\B$ denote a hereditary $C^*$-subalgebra of $\B(\H)$. One of their main results is
an ``asymptotic'' Riesz representation theorem (Theorem \ref{thm:asmriesz}) which gives a
bijective correspondence between certain 
positive asymptotic morphisms (i.e. ``quantizations'') $$\{Q_\hbar\}_{\hbar \in (0,1]} : C_0(X) \to \B$$ and certain Borel
asymptotic spectral measures 
$$\{\Ah\}_{\hbar \in (0,1]} :  (\Sigma_X,\CC_X ) \to  (\B(\H), \B)$$
where $\Sigma_X$ denote the Borel subsets of $X$ and $\CC_X$ denotes the open subsets of $X$ with compact closure. This correspondence is given by an expectation: For all $|\phi\>$ in $\H$ and $f$ in  $C_0(X)$,
$$\<\phi | Q_\hbar(f) | \phi \>= \int_X f(x) \, d \< \phi | \Ah(x) | \phi \>.$$
The associated asymptotic morphism $\{Q_\hbar\} : C_0(X) \to \B$ then allows
one to define an $E$-theory invariant for the asymptotic spectral measure
$\{\Ah\}$  (and hence the quantum system),
$$ \lcl \Ah \rcl =_{\text{def}} \lcl Q_\hbar \rcl \in E(C_0(X), \B)  \cong
E_0(X; \B),$$
in the $E$-homology group of $X$ with coefficients in $\B$. However, we will not discuss such $E$-theory invariants here, preferring to focus more on the relationship between ASMs and operational quantum physics.

The outline of this paper is as follows. In Section II we review the basic Hilbert space model of quantum measurement theory and its relationship with POVMs and $\cs$-algebra quantization theory. Positive asymptotic morphisms associated
to hereditary and nuclear $C^*$-algebras are discussed in Section III. The basic definitions and properties
of asymptotic spectral measures  are reviewed in Section IV.  Asymptotic Riesz representation theorems and some of their consequences are contained in Section V. Examples and applications of ASMs associated to various
aspects of quantum physics are discussed in Section VI, e.g., constructing ASMs from PVMs by quantum noise models, quasiprojectors and semiclassical limits, and unsharp spin measurements of spin-$\frac12$ particles  (including an example from quantum cryptography).

\section{POV-measures and Quantum Theory}

\subsection{Review of Hilbert Space Theory}

We begin by reviewing basic Hilbert space theory, mainly to fix terminology and notation, especially the Dirac bra-ket notation, which is more common in physics than pure math.

A  {\it Hilbert space} \cite{Fo84,Cwy90} is a complex vector space $\H$ equipped with an inner product
\begin{equation*}
\< \cdot | \cdot \> : \H \times \H \to \C
\end{equation*}
such that $\H$ is complete with respect to the norm defined by
\begin{equation*}
\| \phi \| = \sqrt{\<\phi | \phi \>}
\end{equation*}
for all vectors (kets) $| \phi \>$ in $\H$. The inner product has the following properties:
\begin{tightlist}
\item Positivity: $\<\phi | \phi \> \geq 0$ and $\< \phi | \phi \> = 0$ iff $|\phi\> = 0$,
\item Linearity: $\< \phi | \alpha \psi_1 + \beta   \psi_2 \> = \alpha \< \phi | \psi_1 \> + \beta \< \phi | \psi_2 \>$,
\item Skew-symmetry: $\< \phi | \psi \>^* = \< \psi | \phi \>$,
\end{tightlist}
for all $|\phi\>, |\psi\>, |\psi_i\> $ in $\H$ and complex numbers $\alpha, \beta$ in $\C$, where $\alpha^* = a -bi$ denotes the conjugate of the complex number $\alpha = a + bi$ where $i^2 = -1$.

An {\it operator} on $\H$ is a linear transformation
\begin{align*}
& T : \H  \to \H \\
& |\phi\> \mapsto T | \phi \>.
\end{align*} 
The operator $T$ is called {\it bounded} (= {\it continuous}) if the {\it operator norm} of $T$ is finite, i.e., 
\begin{equation}\label{eqn:opnorm}
\| T\| = \sup \{ \| T\phi \| : |\phi \> \in \H,  \| \phi \| = 1 \} < \infty.
\end{equation}  The set of all bounded linear operators $T$ on $\H$ is denoted by $\B(\H)$.
The {\it adjoint} of an operator $T$ in $\B(\H)$ is the (unique) operator $T^\dag : \H \to \H$ in $\B(\H)$ such that
\begin{equation*}
\< \phi | T^\dag | \psi \> = \< \psi | T | \phi \>^*
\end{equation*}
for all $| \phi \>$ and $| \psi \>$ in $\H$.  The adjoint operation $T \mapsto T^\dag$ satisfies the axioms of an {\it involution}: For all $T \in \B(\H)$,
\begin{tightlist}\label{list:inv}
\item Idempotent: $(T^{\dag})^{\dag} = T$;
\item Anti-linear: $(\alpha S + \beta T)^\dag = \alpha^* S^\dag + \beta^* T^\dag$;
\item Anti-multiplicative : $(S \circ T)^\dag = T^\dag \circ S^\dag$
\end{tightlist}
where $(S \circ T)|\phi\>  = S(T |\phi \>)$ denotes the composition of operators $S$ and $T$. It is very closely related to the conjugate transpose of a complex matrix \cite{Fo84}.
A very important result is that the {\it $\cs$-identity} holds: For all $T$ in $\B(\H)$,
\begin{equation}\label{eqn:C*id}
\|T\|^2 = \|T^\dag\|^2 = \|T^\dag \circ T\|^2.
\end{equation}
It follows that $\B(\H)$has the structure of a $\cs$-algebra. (See subsection \ref{subsect:povmquant} below for a discussion of $\cs$-algebras and quantum theory.)

An operator $T$ is called {\it self-adjoint} ({\it Hermitian}) if $T = T^\dag$ or equivalently, if for all $|\phi\>$ and $|\psi\>$ in $\H$,
\begin{equation*}
\<\phi | T | \psi\> = \< \psi | T | \phi \>^*.
\end{equation*}
For example, the identity operator $I$ on $\H$ given by $I |\phi\> = |\phi\>$ is self-adjoint $I = I^\dag$ and has operator norm one $\|I\| = 1$. If $T$ is self-adjoint, then $T$ has real spectrum 
\begin{equation*}
\spec(T) =_\text{def} \{\lambda \in \C : T-\lambda I  \text{ not invertible in } \B(\H) \} \subset \R
\end{equation*}
An operator $T \in \B(\H)$ is called {\it positive}, denoted $T \geq 0$, if 
\begin{equation*}
\< \phi | T | \phi \> \geq 0
\end{equation*}
 for all $| \phi \> $ in $\H$. It follows that $T = T^\dag$ and has nonnegative spectrum $\spec(T) \subset [0, \infty)$.
 A {\it density operator} is a positive operator $T \geq 0$ in $\B(\H)$ of unit trace: $ \tr(T) = 1.$
Note that if $\dim(\H) = \infty$, then there are operators without finite trace \cite{Cwy90}.
There is a partial order $\leq$ on $\B(\H)$ defined by $S \leq T$ on $\B(\H)$ iff  $T- S \geq 0$. There is also an {\it orthogonality} relation $\perp$ given by $S \perp T$ if and only if $\range(S) \perp \range(T)$. This is equivalent to the algebraic condition that
$S^\dag T = T^\dag S = 0$.

\subsection{The von Neumann Model of Quantum Mechanics}

A quantum-mechanical system  \cite{VN32,Ish95,Omn99} is modeled by a Hilbert space $\H$ which forms the {\it state space} of the system. A  ({\it pure}) {\it state} of the system is given by a unit vector $|\phi\>$ in $\H$, i.e., a vector of norm one $\| \phi \| = 1$.  In general, a {\it  state} of the system is given by a density operator $\rho$ on $\H$. For example, a pure state $|\phi\>$ can be represented by the density operator \begin{equation*}
\rho = |\phi\>\<\phi|,
\end{equation*}
where $\rho |\psi \> = |\phi\> \< \phi | \psi\> = \< \phi | \psi\> |\phi\>.$ (Recall that the 'bra' $\< \phi | : \H \to \C$ is the dual of the vector (ket) $|\phi\>$ and is defined by $ |\psi\> \mapsto \<\phi |\psi \>$.) 
More generally, if a quantum system can be in pure states
\begin{equation*}
|\phi_1\>, |\phi_2\>, \cdots, |\phi_n\>,
\end{equation*}
(which need not be orthogonal) each with respective (classical) probabilities 
\begin{equation*}
0 \leq p_1, p_2, \cdots, p_n \leq 1,
\end{equation*}
where $p_1 + p_2 + \cdots + p_n = 1$, then the density operator representing this {\it mixed state} is given by
\begin{equation*}
\rho = p_1 |\phi_1\>\<\phi_1| + p_2 |\phi_2\>\<\phi_2| + \cdots + p_n |\phi_n\>\<\phi_n|.
\end{equation*}
However, this decomposition is not unique.

A ({\it bounded}) {\it observable} of the system is given by a self-adjoint operator $\O = \O^\dag$ in $\B(\H)$. For unbounded observables, we also need to consider self-adjoint operators that are only defined on a dense vector subspace of $\H$, as in the position $\hat{Q} = x$ and momentum operator $\hat{P} = -i \hbar \frac{d}{dx}$ of a free particle moving on a line $\R$ with state space $\H = L^2(\R)$ \cite{Ish95,VN32}. (However, we will not discuss unbounded operators here since the functional analysis is rather technical and the general POVM model of observables incorporates both bounded and unbounded in the same framework)  The measurable values of an observable $\O$ are contained in the spectrum $\spec(\O) \subset \R$, which can be either discrete or continuous. If the system is in state $\rho$ the {\it expected  value} of the observable $\O$ is given by the Born rule:
\begin{equation*}
\< \O \>_\rho = \tr(\rho \circ \O).
\end{equation*}
Note that if $\rho = |\phi\>\<\phi |$ is a pure state, then we recover the usual expectation formula,
\begin{equation*}
\< \O \>_\rho = \tr (|\phi\>\<\phi | \circ \O) =  \tr (\<\phi | \O |\phi\>) = \<\phi | \O |\phi\>,
\end{equation*}
from elementary quantum mechanics. 

Every (closed) quantum mechanical system is equipped with a special observable $\hat{H}$ (which my depend on time $t$) called the {\it Hamiltonian} which measures the total energy and time-evolution of the system. Indeed, if $|\psi(t)\>$ denotes the (pure) state of the system at time $t$ then the dynamical behavior of the system is governed by the {\it Schr\"odinger equation}:
\begin{equation*}
i \hbar \frac{\partial}{\partial t} |\psi(t)\> = \hat{H} |\psi(t)\>.
\end{equation*}
which is a deterministic equation in the state. However, as is well known, quantum theory can only predict the probability distributions of measurements of {\it observables}. To compute the probability of a measurement outcome of an observable while the system is in a certain state, von Neumann generalized the spectral theory of $n \times n$ matrices as follows.

Let $\R$ denote the real numbers with Borel $\sigma$-algebra\footnote{Recall that a $\sigma$-algebra on a set $X$ is a collection $\Sigma$ of subsets of $X$ such that  $X \in \Sigma$, the emptyset $\emptyset \in \Sigma$, and $\Sigma$ is closed under complements $\Delta^c = X \setminus \Delta$ and countable unions $\sqcup_1^\infty \Delta _n$ \cite{Fo84}.} $\Sigma_\R$ generated by the open intervals $(a, b)$.  A {\it projection-valued measure} (PVM or {\it spectral measure}) is a mapping
\begin{align*}
P : \Sigma_\R & \to \B(\H) \\
\Delta & \mapsto P(\Delta)
\end{align*}
such that the following conditions hold:
\begin{tightlist}
\item Normalization: $P(\emptyset) = 0$ and $P(\R) = I$;
\item Projectivity: $P(\Delta)^2 = P(\Delta)^\dag = P(\Delta)$;
\item Orthogonality: $P(\Delta) \perp P(\Delta')$ if $\Delta \cap \Delta' = \emptyset$;
\item $\sigma$-additivity: $\< \phi | P(\sqcup_{n=1}^\infty \Delta _n) | \phi \>= \sum_{n=1}^\infty \< \phi | P(\Delta _n) | \phi \>$
\end{tightlist}
for disjoint measurable sets $\{\Delta _n\}_{n=1}^\infty$ in $\Sigma_\R$ and all $| \phi \>$ in $\H$. The projectivity and orthogonality conditions are equivalent to the condition that $P(\Delta) \geq 0$ and 
\begin{equation}\label{cond:proj}
P(\Delta \cap \Delta') = P(\Delta)P(\Delta')
\end{equation}
for all (Borel) subsets $\Delta$ and $\Delta'$ of $\R$.

\begin{ex}\label{ex:finsys} Consider a finite $n$-level quantum system with state space $\H = \C^n$.
Let $\O$ be a observable on $\H$ with nondegenerate spectrum (i.e.,  eigenvalues)
\begin{equation*}
\spec(\O) = \{\lambda_1, \lambda_2, \cdots, \lambda_n\} \subset \R.
\end{equation*}
Let $P_k  : \C^n \to E_k$ denote the orthogonal projection onto the (one-dimensional) $k$th-eigenspace $E_k$, that is,
\begin{equation*}
E_k = \{ |\phi\> \in \C^n :  \O |\phi\>  = \lambda_k |\phi\> \}.
\end{equation*}
It follows that we can write $P_k = |k\> \< k |$, where $|k\> \in E_k$ is a unit vector.
The spectral decomposition of $\O$ is then given by
\begin{equation}
\O = \sum_{k = 1}^n \lambda_k P_k = \sum_{k = 1}^n \lambda_k  |k\> \< k |
\end{equation}
and the associated spectral measure on $\R$  is given by
\begin{equation*}
P(\Delta) = \sum_{\lambda_k \in \Delta} P_k = \sum_{\lambda_k \in \Delta}  |k\> \< k |
\end{equation*}
for all Borel subsets $\Delta \subseteq \R$. The orthogonality condition above corresponds to the fact that eigenspaces of self-adjoint operators are orthogonal:
\begin{equation*}
E_k \perp E_j \text{ if and only if } k \neq j.
\end{equation*}
The normalization condition above corresponds to the {\it resolution of the identity} formula
\begin{equation*}
 \sum_{k=1}^n |k\> \< k | =  \sum_{k=1}^n P_k = P(\R)= I.
\end{equation*}
\smallskip
\end{ex}

The fundamental result in the von Neumann formulation of quantum theory is the following Spectral Theorem of Hilbert.

\begin{thm}[Spectral Theorem]
There is a one-one correspondence between Borel spectral measures
$$P : \Sigma_\R \to \B(\H)$$ on $\R$ and
self-adjoint operators $\O$ on the associated Hilbert space $\H$. This correspondence
is given by the formula:
$$\< \phi | \O | \phi \> = \int_{-\infty}^\infty \lambda \, d \< \phi | P(\lambda) | \phi \>.$$
for all $|\phi\>$ in $\H$.
\end{thm}

If the self-adjoint operator $\O$ has a discrete (finite) spectrum (as in Example \ref{ex:finsys}), the (weak) integral above degenerates to the familiar sum:
$$\< \phi | \O | \phi \> = \sum_{k=1}^n \lambda_k  \< \phi | P_k | \phi \> = \sum_{k=1}^n \lambda_k  |\<k|\phi\>|^2.$$
The physical (operational)  interpretation of the map 
\begin{equation*}
\Delta \mapsto P(\Delta)
\end{equation*}
is the probability $\Prob(\Delta | \rho)$ that a measurement of the observable $\O$, while the quantum system is in a state represented by density operator $\rho$, will yield a value in the subset $\Delta \subset \R$ of is given by the Born probability rule:
\begin{equation*}
 \Prob(\Delta | \rho) = \tr(\rho  \circ P(\Delta)).
 \end{equation*}
 Since $0 \leq P(\Delta) \leq I$ it follows that
 \begin{equation*}
 0 \leq \tr(\rho \circ P(\Delta) ) \leq \tr(\rho \circ I) = \tr(\rho) = 1
 \end{equation*}
 which shows that we obtain probabilities in $[0,1]$.

\subsection{POVMs and Generalized Measurements}
 
 In order to solve unsharpness issues with measurements of certain quantum systems and to deal with foundational
problems, the concept of observable was enlarged  to include more general kinds of operator-valued measures. See the discussions (and references) in the monographs \cite{Sch96,BGL95} for a thorough exposition of POVMs in the foundational and operational aspects of quantum physics. See Brandt \cite{Bran99} for a short history of POVMs in quantum theory and an application to photonic qubits in quantum information processing. See Berberian \cite{Ber66} for the basic integration theory of POVMs.

Let $\H$ be the Hilbert state space of a quantum-mechanical system. Let $X$ be a set equipped with a $\sigma$-algebra $\Sigma$ of measurable subsets of $X$. A {\it positive operator-valued measure} 
(or POVM) on the measurable space $(X, \Sigma)$ is a mapping 
\begin{equation*}
A : \Sigma \to \B(\H)
\end{equation*}
which satisfies the following properties:
\begin{tightlist}
\item Nullity: $A(\emptyset) = 0$,
\item Positivity: $A(\Delta) \geq 0$ for all $\Delta$ in $\Sigma$,
\item $\sigma$-Additivity: $\< \phi | A(\sqcup_{n=1}^\infty \Delta _n) | \phi \>= \sum_{n=1}^\infty \< \phi | A(\Delta _n) | \phi \>$
\end{tightlist}
 for disjoint measurable sets $\{\Delta _n\}_{n=1}^\infty$ in $\Sigma$ and all $| \phi \>$ in $\H$. Note that positivity and additivity imply that
\begin{equation*}
0 \leq A(\Delta) \leq A(X) < \infty
\end{equation*} and so $\|A(\Delta)\| \leq \|A(X) \| < \infty$  for all $\Delta \in \Sigma$.
We will say that $A$ is {\it normalized} if  $A(X) = I$. If each $A(\Delta )$
is a projection in $\B(\H)$, i.e.,  
\begin{equation*}
A(\Delta )^2 = A(\Delta)^\dag = A(\Delta ),
\end{equation*}
then we call $A$ a {\it projection-valued measure} (PVM or {\it spectral
measure}) on $X$. This is equivalent to the condition that:
For all $ \Delta, \Delta' \in \Sigma$,
\begin{equation}\label{cond:projA}
A(\Delta \cap \Delta') = A(\Delta)A(\Delta')
\end{equation}
which is comparable to condition (\ref{cond:proj}) above.

A {\it generalized (unsharp) observable} can be modelled by a normalized POV-measure $A$ on the classical phase (or configuration) space $X$ of a quantum system.
The physical interpretation of the map $\Delta \mapsto
A(\Delta)$ is the probability that the physical system,  in a state represented
by density operator $\rho$, is
localized in the subset $\Delta$ of the  space $X$ is given by the number 
\begin{equation*}
 \Prob(\Delta|\rho) = \tr(\rho  \circ A(\Delta)) = \tr \Big( \int_\Delta \rho \ dA \Big).
 \end{equation*}
If $X = \R$ then the mean or vacuum expectation value of the associated (unsharp) quantum observable is then
computed by the formula
\begin{equation*}
\< A \>_\rho  = \tr \Big( \int_{-\infty}^{\infty} \lambda
\rho(\lambda) \ dA(\lambda) \Big).
\end{equation*}
where $p(\lambda)$ is the probability density of the state $\rho$.

\begin{ex}\label{ex:spinhalf}
Recall that pure spin-$\frac12$ systems are represented by the Hilbert state space $\H = \C^2$
\cite{BGL95,Sch96}. We then have $\B(\H) \cong M_2(\C)$. The Pauli spin operators
$\sigma_1, \sigma_2, \sigma_3$
are the $2 \times 2$ matrices 
$$\sigma_1 = \begin{pmatrix} 0 & 1 \\ 1 & 0 \end{pmatrix}, \sigma_2 = \begin{pmatrix} 0 & -i \\
i & 0 \end{pmatrix}, \sigma_3 = \begin{pmatrix} 1 & 0 \\ 0 & -1 \end{pmatrix} $$
which satisfy the relations:
\begin{tightlist}
\item $\sigma_i^\dag = \sigma_i, \sigma_i^2 = I$
\item $\sigma_i \sigma_j = - \sigma_i \sigma_j$ for $i \neq j$
\item $\sigma_i \sigma_j = i \epsilon_{ijk}\sigma_k$ for $i \neq j$
\end{tightlist}
where $I$ denotes the identity operator. A density  operator (or state) on $\H$
is then a positive matrix $\rho \geq 0$ with trace one.
A fundamental result in the theory is the following.

\begin{lem} Any density operator $\rho$ on $\H$ can be written
uniquely in the form
$$\rho = \rho(\vec x) = \frac12 (I + \vec x \cdot \vec \sigma), \quad \vec x =
(x_1, x_2, x_3)  \in
\R^3, \quad \|\vec x \| \leq 1,$$
where $\vec x \cdot \vec \sigma = x_1 \sigma_1 + x_2 \sigma_2 + x_3 \sigma_3$
and $\|\vec x\|^2 = x_1^2 + x_2^2 + x_3^2$.
Moreover, $\rho$ is a one-dimensional projection
iff $\vec x$ is a unit vector $\|\vec x \| =1$.
\end{lem}

Let $X_{\text{spin}} =  \{-\frac12, +\frac12\}$ denote the set of measurements (i.e., meter readings) corresponding to ``spin-down'' ($-\frac12$) and ``spin-up'' ($+\frac12$), respectively. A POVM $A$ on $X_{\text{spin}}$ is then uniquely characterized by two positive matrices
$ A^+, A^- \geq 0$, where $A^- = A(\{-\frac12\})$ and $A^+ = A(\{+\frac12\})$, and $A^+ + A^- = I$.

\begin{dfn}\label{dfn:spinpovm} A {\it spin} POVM on $X_{\text{spin}} = \{-\frac12,
+\frac12\}$ is a normalized POVM $A = \{A^+, A^-\}$ such that
$\tr(A^\pm) = 1$. Thus, each  $A^\pm \geq 0$ is a density operator and they sum to $I$.
\end{dfn}

Let $B^3 = \{\vec x \in \R^3 : \|\vec x\| \leq 1\}$ denote the closed unit
ball in Euclidean $3$-space  $\R^3$. Let $S^2 = \partial B^3$ denote the unit sphere.
For each $\vec x \in B^3$ we obtain 
a spin POVM $A_x$ on $X_{\text{spin}}$ by defining
\begin{equation}\label{eqn:spinpovm}
A_x^\pm = \rho(\pm \vec x) = \frac12(I \pm \vec x \cdot \vec \sigma)
\end{equation}
which determines an ``unsharp'' spin observable.
Let $\lambda = \lambda(\vec x) = \|\vec x\|$ and define the quantities
$$r_x = \frac{1+ \lambda(\vec x)}{2} > \frac12,  \quad u_x =
\frac{1-\lambda(\vec x)}{2} < \frac12.$$
The quantity $r_x$ is called the degree of reality and $u_x$ is the degree of
unsharpness of the unsharp observable $A_x$ \cite{BGL95,RK99}.

\begin{lem} There is a bijective correspondence between spin POVMs $A
= \{A^+, A^-\}$ and points
$\vec x \in B^3$ in the closed unit ball of $\R^3$ given by (\ref{eqn:spinpovm}).
Moreover, $A$ is a spectral measure if and only if  $\vec x \in S^2$ is a unit
vector.
\end{lem}

\end{ex}

\subsection{POVMs and $\cs$-algebra Quantization}\label{subsect:povmquant}

Recall that if $X$ is the classical phase/configuration space (or space-time) of a classical (relativistic) physical system, then classical observables are given by (continuous or measurable) real-valued functions $f : X \to \R$. The value of the observable $f$ when the system is in state $s$ in $X$ is given by the real number $f(s)$. This naturally leads us to consider various algebras of functions on $X$.

If $X$ is a locally compact Hausdorff topological space we let $C_b(X)$ denote
the (complex) algebra of all continuous and bounded functions $f : X \to \C$, with algebraic operations inherited pointwise from $\C$. This algebra forms a Banach (i.e complete normed) $*$-algebra with the supremum norm
\begin{equation*}
\|f\| = \sup \{ |f(x)| : x \in X \}\
\end{equation*}
and involution given by pointwise conjugation
\begin{equation*}
f^\dag(x) = f(x)^*
\end{equation*}
for all $f$ in $C_b(X)$ and $x$ in $X$. We let $C_0(X)$ denote the subalgebra consisting of all $f$ in $C_b(X)$ which ``vanish at infinity'' on $X$, i.e.,
\begin{equation*}
\lim_{x \to \infty} |f(x)| = 0.
\end{equation*}
Note that $C_0(X)$ is the completion in $C_b(X)$ of the non-closed ideal $C_c(X)$ of functions with compact support. All of these are subalgebras of the algebra $B_b(X)$ of bounded measurable functions $f :X \to \C$ with the same norm and involution given above. (However, if $X$ is second countable, then $C_0(X)$ is separable, but both $C_b(X)$ and $B_b(X)$ are, in general, nonseparable.) The algebras $C_0(X) \subset C_b(X) \subset B_b(X)$ also have the important structure of being $\cs$-algebras.

\begin{dfn} A {\it $\cs$-algebra} \cite{Mur90} is a complex normed algebra $A$ with norm $a \mapsto \|a\|$ and involution $a \mapsto a^\dag$ (see \ref{list:inv})
such that for all $a$ in $A$ the $\cs$-identity holds: For all $a$ in $A$,
\begin{equation*}
\|a^\dag a\| = \|a\|^2
\end{equation*}
and $A$ is complete with respect to the norm. An element $a$ in $A$ is called {\it self-adjoint} if $a^\dag = a$ and is called {\it positive} ($a \geq 0$) if $a = b^\dag b$ for some element $b$ in $A$. We then define $a \leq b$ if $b - a \geq 0$.
\end{dfn}

For example, $\B(\H)$ is a $\cs$-algebra with multiplication given by operator composition $ST = S \circ T$,  the operator norm $\|T\|$ given by (\ref{eqn:opnorm}), and involution given by the adjoint operation $T \mapsto T^\dag$. Thus, the self-adjoint part of the $\cs$-algebra $\B(\H)$ can represent bounded {\it quantum} observables. Similarly, the self-adjoint part of the $\cs$-algebra $C_b(X)$ can represent bounded and continuously-varying {\it classical} observables. However, in practical experiments, states of classical physical systems occupy bounded (compact) regions of the space-time or phase space $X$, so it is generally easier to work with the $\cs$-algebra $C_0(X)$, since it is usually separable.  For the importance of general $\cs$-algebras to quantum theory, see the books by Emch \cite{Em72} and Landsman \cite{Land98}.

Historically, the  ``quantization'' of a classical physical system meant a (usually ad hoc) method of associating to a classical observable $f$ a corresponding quantum observable $Q(f)$, which obeyed certain rules. This leads to the following general definition.

\begin{dfn}
A {\bf general quantization} of a classical space $X$ on a Hilbert space $\H$ is a linear
map 
\begin{equation*}
Q : C_0(X) \to \B(\H)
\end{equation*}
which is also {\it positive}. That is, $Q(f) \geq 0$ for all $f \geq 0$.
If $X$ is compact, we require that $Q(1) = I$. If
$X$ is a non-compact space, we require that $Q$ should have a unital extension
\begin{equation*}
Q^+ : C_0(X)^+ = C(X^+) \to \B(\H)
\end{equation*}
which is a positive linear map, where $X^+ = X \cup \{\infty\}$ denotes the one-point compactification of $X$ \cite{Mun75}.
\end{dfn}

The positivity condition implies, in particular,  that real-valued functions $f = f^\dag$ are mapped to self-adjoint operators $Q(f)^\dag = Q(f)$. If $Q : C_0(X) \to \B(\H)$ is a $*$-homomorphism, i.e., a linear map such that $Q(f g) = Q(f) \circ Q(g)$ and $Q(f^\dag) = Q(f)^\dag$ then $Q$ is positive since $Q(f f^\dag) = Q(f) \circ Q(f)^\dag$.
Another reason for the importance of positive linear maps in quantum theory is the following
generalized Riesz representation theorem for the dual of $C_0(X)$. (Compare
Proposition 1.4.8 \cite{Land98} and Theorem 19 \cite{Ber66}). Let $\Sigma_X$ be the Borel $\sigma$-algebra generated by the open subsets of the locally compact topological space $X$.

\begin{thm}\label{thm:rieszrep} There is a one-one
correspondence between positive linear maps  $Q : C_0(X) \to \B(\H)$ and
POV-measures
$A : \Sigma_X \to \B(\H)$, given, for all $|\phi \>$ in $\H$ by
\begin{equation}\label{eqn:weakopint}
\< \phi | Q(f) | \phi \> = \int_X f(x) \, d\< \phi | A(x)  |\phi \>)
\end{equation}
The map $Q$ is a general quantization if and only if $A$ is a normalized POVM.
Moreover, $Q$ is a $*$-homomorphism if and only if $A$ is a spectral measure
(PVM).
\end{thm}

The (weak) integral equation  (\ref{eqn:weakopint}) above is then usually abbreviated
\begin{equation*}
Q(f)  = \int_X f(x) \, dA(x)
\end{equation*}
for all $f \in C_0(X)$. The map $Q$ then easily extends to $$Q: B_b(X) \to \B(\H)$$ and satisfies (Theorem 10
\cite{Ber66}): For all $f \in
C_0(X) \subset B_b(X)$,
\begin{equation*}
\|Q(f)\| = \left \| \int_X f(x) \, dA(x) \right \| \leq 2 \|f\| \|A(X)\|.
\end{equation*}
Note that by the Naimark Extension Theorem \cite{RS90}, every
POVM $A$ is the compression of a spectral measure $P$ defined on a
minimal extension $\H' \supset \H$. That is, $$A(\Delta) = Q \circ P(\Delta) \circ Q,$$ where
$Q : \H' \to \H$ is the orthogonal
projection. One could then try to compute the integrals
$\int_X f(x) \ dA(x)$ by computing $\int_X f(x) \ dP(x)$ on $\H$ and then
projecting back down to $\H$. There are two
problems with this \cite{Sch96}. The first is that $\H'$ could have no physical
meaning,
thus making the analysis unsatisfying  to
the physicist. Also,  the integration process may not commute with the
projection process (e.g., when the associated observable is unbounded).

Let $\A$ be a $*$-subalgebra of a $C^*$-algebra $\B$. Recall that $\A$ is said
to be {\bf hereditary} \cite{Mur90} if $0 \leq b \leq a$ and
$a \in \A$ implies that $b \in \A$. Every (closed two-sided $*$-invariant) ideal
in a $C^*$-algebra is a hereditary $*$-subalgebra. 
In particular, if $\H$ is a Hilbert space, the ideal of compact operators
$\K(\H)$ is a hereditary $C^*$-subalgebra of the
$C^*$-algebra of bounded operators $\B(\H)$. An important (non-closed)
hereditary $*$-subalgebra
for quantum theory is the (non-closed) ideal $\B_1(\H) \subset \K(\H)$ of
trace-class
operators: 
\begin{equation*}
\B_1(\H) = \{ \rho \in \K(\H) : \tr |\rho| < \infty \}.
\end{equation*}
We then have that $\K(\H)^* = \B_1(\H)$ by the dual pairing 
\begin{equation*}
(\rho, T) = \tr(\rho \circ T)
\end{equation*} 
where $\rho \in \B_1(\H)$ and $T \in \K(\H).$

If $X$ is a locally compact space, then the ideal $C_0(X)$ is a hereditary
$C^*$-subalgebra of the 
$C^*$-algebra $C_b(X)$ of  continuous bounded complex-valued functions on $X$.
Also, the (non-closed) ideal $C_c(X)$ 
of compactly supported functions is a (non-closed) hereditary $*$-subalgebra of
$C_b(X)$. However, in general, $C_\delta(X)$, for $\delta = $ c, $0$, b,
is not a hereditary subalgebra of the $C^*$-algebra $B_b(X)$ of bounded Borel
functions on $X$.

\begin{dfn} Let $\A_1 \subset \A$ and $\B_1 \subset \B$ be 
subalgebras of the $C^*$-algebras $\A$ and $\B$. If $Q : \A \to \B$ is a
linear map such that $Q(\A_1) \subset \B_1$, we will denote this
by
\begin{equation*}
Q :   (\A,\A_1) \to  (\B,\B_1).
\end{equation*}
\end{dfn}

Since a positive linear map is automatically continuous, we have the following.

\begin{lem} Let $\A_1 \subset \A$ and $\B_1 \subset \B$ be non-closed
$*$-subalgebras.
Every positive linear map $Q : (\A,\A_1) \to  (\B,\B_1)$
also satisfies 
\begin{equation*}
Q : (\A,\overline {\A_1}) \to  (\B, \overline{\B_1}),
\end{equation*}
where $\overline{\A_1}$ denotes the closure of $\A_1 \subset \A$ (similarly for
$\overline{\B_1}$).
\end{lem} 

\section{Asymptotic Morphisms of $\cs$-algebras}

Let $\A$ and $\B$ be $C^*$-algebras. Recall that a linear map $Q : \A
\to \B$
is called {\it positive} \cite{Mur90} if $Q(f) \geq 0$ for all $f \geq 0$ in $\A$. It is
called {\it completely positive} if every inflation  to $n \times n$ matrices $M_n(Q) : M_n(\A) \to
M_n(\B)$ is also 
positive. Every $*$-homomorphism from $\A$ to $\B$ is clearly completely
positive. The following definition
interpolates between (completely) positive linear maps and $*$-homomorphisms.

\begin{dfn}\label{def:cpam} A ({\it completely}) {\it positive asymptotic
morphism} from $\A$ to $\B$ is a
family of maps
\begin{equation*}
\{Q_\hbar\}_{\hbar \in (0,1]} : \A \to \B
\end{equation*}
parameterized by  $\hbar \in (0,1]$ such that the following hold:
\begin{enumerate}
\item Each $Q_\hbar : \A \to \B$ is a (completely) positive linear map; 
\item The map $(0,1] \to \B: \hbar \to Q_\hbar(f)$ is continuous for each
$f \in \A$;
\item For all $f, g \in \A$ we have 
\begin{equation*}
\lim_{\hbar \to 0} \| Q_\hbar(fg) - Q_\hbar(f)Q_\hbar(g)\| = 0.
\end{equation*}
\end{enumerate}
\end{dfn} 
For the basic theory of asymptotic morphisms see the books
\cite{GHT00,Con94,Bla98} and papers \cite{CH90,Gue99}. For the importance
of positive asymptotic morphisms to $C^*$-algebra $K$-theory see \cite{HLT99}.
Note that any $*$-homomorphism
$Q : \A \to \B$ determines the constant completely positive asymptotic morphism
$\{Q_\hbar\} : \A \to \B$
defined by $Q_\hbar = Q$ for all $\hbar > 0$. Also, it follows that for any
$f\in \A$,
a mild boundedness condition \cite{Bla98} holds,
\begin{equation*}
\limsup_{\hbar \to 0} \| Q_\hbar(f) \| \leq \|f\|.
\end{equation*}

\begin{rmk}\label{rem:parm} Asymptotic morphisms, in the $E$-theory literature, are usually
parameterized by $t \in [1, \infty)$. We chose to
use the equivalent parameterization $\hbar = 1/t \in (0,1]$ to make the
connections to quantum physics more transparent. Note that
other authors have used different parameter spaces, including discrete ones
\cite{Lor96,Th03}. The results in this paper translate verbatim
to these parameter spaces, and condition (2) is obviously irrelevant in the
discrete case.
\end{rmk}

\begin{dfn} Two asymptotic morphisms 
\begin{equation*}
\{Q_\hbar\}, \{Q_\hbar'\} : \A \to \B
\end{equation*}
 are called {\it equivalent}
if for all $f \in \A$ we have that 
\begin{equation*}
\lim_{\hbar \to 0} \| Q_\hbar(f) - Q_\hbar'(f) \| = 0.
\end{equation*}
 We let $\lcl \A, \B\rcl_{a(cp)}$ denote the collection of all asymptotic equivalence
classes of (completely positive) asymptotic morphisms from $\A$ to $\B$.
\end{dfn}

A $C^*$-algebra $\A$ is called {\it nuclear} \cite{Mur90} if the identity map $\id : \A
\to \A$ can be approximated pointwise in norm
by completely positive finite rank contractions. This is equivalent to the
condition that there is a 
unique $C^*$-tensor product $\A \tensor \B$ for any $C^*$-algebra $\B$. If $\H$
is a separable Hilbert space, the
$C^*$-algebra $\K(\H)$ of compact operators on $\H$ is nuclear. However, $\B(\H)$ is not nuclear. If $X$ is a
locally compact space, then the $C^*$-algebras
$C_0(X)$ and $C_b(X)$ are nuclear.

If $\A \cong C(X)$ is unital and commutative, then every positive
linear map $Q : \A \to \B$ is completely positive by Stinespring's Theorem.
The following result is a consequence of the completely positive lifting theorem
of Choi and Effros \cite{CE76} for nuclear $C^*$-algebras. 
(See also 25.1.5 of Blackadar \cite{Bla98} for a discussion.)

\begin{lem} Let $\A$ be a nuclear $C^*$-algebra. Every asymptotic
morphism from $\A$ to any $C^*$-algebra $\B$ is
equivalent to a completely positive asymptotic morphism. That is, there is a
bijection of sets
$\lcl \A, \B \rcl_a \cong \lcl \A, \B \rcl_{acp}.$
\end{lem}

\section{Asymptotic Spectral Measures}

In this section we review the basic definitions and properties of asymptotic
spectral measures.  For full details, please see the original paper by Martinez-Trout \cite{MT02}.

Let $\H$ be a fixed Hilbert space. Let $X$ be a set equipped with a $\sigma$-algebra $\Sigma$ of measurable sets. Let $\E \subset \Sigma$ denote a fixed collection of measurable subsets. The following definition is intended to interpolate between the definition of a spectral measure (PVM) and a POV-measure, just as the definition of a positive asymptotic morphism interpolates between a $*$-homomorphism and a positive linear map. The key is the asymptotic version of the projectivity conditions (\ref{cond:proj},\ref{cond:projA}), which mimics the asymptotic multiplicity condition in Definition 3.1(3) of a positive asymptotic morphism.

\begin{dfn}\label{def:asm}
A {\bf asymptotic spectral  measure} (ASM) on $(X,\Sigma, \E)$ is a family of
maps 
\begin{equation*}
\{\Ah\}_{\hbar \in (0,1]} : \Sigma \to \B(\H)
\end{equation*}
parameterized by $\hbar \in (0,1]$ such that the following hold:
\begin{enumerate}
\item Each $\Ah : \Sigma \to \B(\H)$ is a POVM;
\item $\limsup_{\hbar \to 0} \|\Ah(X)\| \leq 1$;
\item The map $(0,1] \to \B(\H) : \hbar \to \Ah(\Delta)$ is continuous for
each $\Delta \in \E$;
\item For each $\Delta$, $\Delta'$ in $\E$ we have that
\begin{equation*}
\lim_{\hbar \to 0} \|  \Ah(\Delta \cap \Delta') - \Ah(\Delta) \Ah(\Delta') \| = 0.
\end{equation*}
\end{enumerate}
The triple $(X, \Sigma, \E)$ will be called an {\it asymptotic measure space}.
The family $\E$ is called the {\it asymptotic carrier} of $\{\Ah\}$.
Condition (4) will be called {\it asymptotic projectivity} (or {\it
quasiprojectivity}) and generalizes the projectivity conditions (\ref{cond:proj}, \ref{cond:projA}) of 
spectral measures. It is motivated by the quantum theory notion of
quasiprojectors, as discussed in Example \ref{ex:semi}. If $\E = \Sigma$ then we will call
$\{\Ah\}$ a {\it full}
ASM on $(X, \Sigma)$. If each $\Ah$ is normalized, i.e., $\Ah(X) =I$,
then we will say that $\{\Ah\}$ is {\it normalized}. The mild
boundedness condition (2) is then redundant. (Also see Remark \ref{rem:parm}.)
\end{dfn}

A spectral measure (PVM) $P : \Sigma \to \B(\H)$ determines the ``constant'' full
asymptotic spectral measure $\{\Ah\}$ by the assignment $\Ah(\Delta) = P(\Delta)$
for all $\hbar$ and $\Delta$ in $\Sigma$. Also, any continuous family $\{P_\hbar\}$ of spectral measures
(in the sense of (3)) determine an ASM on $(X,\Sigma, \E)$.
See \cite{CHM00} for an application of smooth families of spectral measures to the
Quantum Hall Effect. 

\begin{dfn} Two asymptotic spectral measures 
\begin{equation*}
\{\Ah\}, \{\Bh\} : \Sigma \to \B(\H)
\end{equation*}
on $(X, \Sigma, \E)$ are said to be
{\it (asymptotically) equivalent} if for each measurable set $\Delta \in \E$,
\begin{equation*}
\lim_{\hbar \to 0} \| \Ah(\Delta) - \Bh(\Delta) \| = 0.
\end{equation*}
This will be denoted $\{\Ah\} \sim_\E \{\Bh\}$. If this holds for $\E = \Sigma$
we will
call them {\it fully equivalent}.
\end{dfn}

From now on, we let $X$ denote a locally compact Hausdorff topological space
with
Borel $\sigma$-algebra $\Sigma_X$. We will assume that $\E = \CC_X$ denotes the
collection of all open
subsets $U$ of $X$ with compact closure $\bar{U}$, i.e., the pre-compact open subsets.

\begin{dfn} Let $\B \subset \B(\H)$ be a hereditary
$*$-subalgebra.
A Borel POVM $A : \Sigma_X \to \B(\H)$
is called {\it locally $\B$-valued} if $A(U) \in \B$ for all pre-compact open subsets
$U \in \CC_X$. This will be denoted by
\begin{equation*}
A : (\Sigma_X, \CC_X) \to (\B(\H), \B).
\end{equation*}
A family of Borel POV-measures $\{\Ah\}$ on $X$ will be called
{\it locally $\B$-valued} if each
POVM $\Ah$ is locally $\B$-valued and will be denoted
$\{\Ah\} :   (\Sigma_X,\CC_X) \to  (\B(\H),\B).$
We will use the term {\it locally compact-valued} for locally $\K(\H)$-valued.
If $\B =
\B_1(\H) \subset \K(\H)$ is the trace-class operators,
then we will say that $\{\Ah\}$ has {\it locally compact trace}.
\end{dfn}

We will let $\( X, \B\)$ denote the set of all equivalence classes 
of locally $\B$-valued Borel asymptotic spectral measures on $(X,\Sigma_X,
\CC_X)$. The equivalence class
of $\{\Ah\}$ will be denoted $\(\Ah\) \in \(X, \B\)$.

Given a Borel POV-measure $A$ on $X$, the {\it cospectrum} \cite{Ber66} of $A$ is defined as
the set
\begin{equation*}
\cospec(A) = \bigcup \{U \subset X: U \text{ is open and } A(U) = 0\}.
\end{equation*}
The {\it spectrum} of $A$ is the complement 
\begin{equation*}
\spec(A) = X \backslash \cospec(A).
\end{equation*}
The following definition is adapted from Berberian \cite{Ber66}.

\begin{dfn} A POVM $A$ on $X$ will be said to have {\it compact
support} if the spectrum of $A$ is a compact
subset of $X$. An ASM $\{\Ah\}$ on $X$ will be said
to have {\it compact support} if there is 
a compact subset $K$ of $X$ such that $\spec(\Ah) \subseteq K$ for all $\hbar >
0$.
\end{dfn}

The relationship among these compactness notions is contained in the
following (which is Proposition 3.5 \cite{MT02}.)

\begin{prop} Let $X$ be second countable. Let $A$
be a Borel POVM on $X$ with compact support. Let $\B$ be the 
hereditary subalgebra of $\B(\H)$ generated by $A(\spec(A))$. Then $A$ is a
locally $\B$-valued POVM, i.e., 
\begin{equation*}
A : (\Sigma_X,\CC_X) \to  (\B(\H),\B).
\end{equation*}
\end{prop}

\section{Asymptotic Riesz Representation Theorems}

Throughout this section, we let $X$ denote a locally compact Hausdorff space
with Borel $\sigma$-algebra $\Sigma_X$. Let
$\CC_X \subset \Sigma_X$ denote the collection of all pre-compact open subsets
of $X$. Thus, we are considering the asymptotic measure space $(X, \Sigma_X, \CC_X)$. And we let $\B \subset \B(\H)$
denote a hereditary $\cs$-subalgebra of the bounded operators on a fixed Hilbert
space $\H$. For proofs of the results in this section, see Section 4 of the original paper by Martinez-Trout \cite{MT02}.

\begin{lem} There is a bijective correspondence between
locally
$\B$-valued Borel POVMs $A : (\Sigma_X,\CC_X) \to  (\B(\H),\B )$
and positive linear maps $Q : C_0(X) \to \B$. This correspondence
is given
by \begin{equation*}
Q(f) = \int_X f(x) \, dA(x)
\end{equation*}
where we interpret this integral as in Equation (\ref{eqn:weakopint}).
\end{lem}

Define $B_0(X)$ to be the $C^*$-subalgebra of $B_b(X)$ generated by $\{\chi_U :
U \in \CC_X\}$ where $\chi_U$ denotes the characteristic function of $U \subseteq X$.
If $X$ is also $\sigma$-compact, a paracompactness argument then shows that
$C_0(X) \subset B_0(X)$ as a closed
(but not necessarily hereditary) $*$-subalgebra. (Recall that if $f \in C_c(X)$
is compactly supported, then
$\Interior(\supp(f)) \in \CC_X$.) The following is our main result, which is the asymptotic version
of Theorem \ref{thm:rieszrep}

\begin{thm}\label{thm:asmriesz} If $X$ is $\sigma$-compact, there is a bijective
correspondence between
positive asymptotic morphisms
\begin{equation*}
\{Q_\hbar\} :( B_0(X),  C_0(X)) \to  (\B(\H), \B)
\end{equation*}
and locally $\B$-valued Borel asymptotic spectral measures
$$\{\Ah\} : ( \Sigma_X, \CC_X ) \to (\B(\H), \B).$$
This correspondence is given by
\begin{equation*}
Q_\hbar(f) = \int_X f(x) \, d\Ah(x).
\end{equation*}
\end{thm}
\smallskip

\begin{cor} With the above hypotheses, equivalent Borel asymptotic
spectral measures correspond to equivalent positive asymptotic morphisms. That is,
there is
a well-defined map 
\begin{equation*} 
\(X, \B\) \to \lcl C_0(X), \B \rcl_{acp}
\end{equation*}
 which maps $\(\Ah\) \mapsto \lcl Q_\h \rcl_{acp}$.
\end{cor}

Let $C_\delta(X)$ denote a unital $C^*$-subalgebra of $C_b(X)$ such that $C_0(X)
\ideal C_\delta(X)$.
By the Gelfand-Naimark Theorem \cite{GN43}, $C_\delta(X) \cong C(\delta X)$ for
some `continuous' compactification $\delta X \supseteq X$.

\begin{cor} Let $\I \ideal \B(\H)$ be an ideal. Every locally
$\I$-valued
full Borel asymptotic spectral measure $\{\Ah\}$
on $X$ determines a canonical relative asymptotic morphism (in the sense of
Guentner \cite{Gue99})
$$\{Q_\hbar\} : (C_\delta(X), C_0(X)) \to (\B(\H), \I) $$
for any continuous compactification $\delta
X$ of $X$.
\end{cor}

\begin{dfn} A family $\{\Ah\}_{\hbar > 0} : \Sigma \to \B(\H)$
of Borel POV-measures on $X$ will be called a
{\it $C_\delta$-asymptotic spectral measure} if the family of maps $\{Q_\hbar\}$
defined by equation (4.1)
determines an asymptotic morphism $\{Q_\hbar\} : C_\delta(X) \to \B(\H)$.
\end{dfn}

The following proposition is then easy to prove using Theorem \ref{thm:rieszrep} and the
results above.

\begin{prop} There is a one-one correspondence between
locally $\B$-valued $C_\delta$-asymptotic spectral measures
$$\{\Ah\} : ( \Sigma_X, \CC_X ) \to (\B(\H), \B)$$ and positive
asymptotic morphisms $$\{Q_\hbar\} :  ( C_\delta(X), C_0(X)) \to  
(\B(\H), \B).$$
\end{prop}

\section{Examples and Applications}
 
 In this section, we look at examples and applications of asymptotic spectral measures to various
 aspects of quantum theory. See Section 5 of Martinez-Trout \cite{MT02} for full details, proofs, and more examples.

\subsection{ASMs and Quantum Noise Models }\label{ex:noise}

A general method for constructing asymptotic spectral measures from
spectral measures (on a possibly different measure space) can be given
by adapting a convolution technique used to model noise and uncertainty in
quantum measuring devices. 
See Section II.2.3 of Busch et al \cite{BGL95} for the relevant background
material. See Beggs \cite{Begg00} for a related method of obtaining asymptotic
morphisms by an integration technique involving spectral measures.

Let $(X_1, \Sigma_1)$ and $(X_2, \Sigma_2)$ be measure spaces. Let $\E_2 \subset
\Sigma_2$.
Consider a family of maps
$$\{p_\hbar\} : \Sigma_2 \times X_1 \to [0,1]$$
such that the following conditions hold:
\begin{enumerate}
\item For every $\omega \in X_1$, $\Delta \mapsto p_\hbar(\Delta, \omega)$
is a probability measure on $X_2$;
\item  For each $\Delta \in \E_2$, the map $\hbar \to p_\hbar(\Delta,
\cdot)$ is continuous $[0,1) \to B_b(X_1)$; 
\item For every $\Delta_1, \Delta_2 \in \E_2$, $$\lim_{\hbar \to 0} \|
p_\hbar(\Delta_1, \cdot) p_\hbar(\Delta_2, \cdot) - p_\hbar(\Delta_1 \cap
\Delta_2, \cdot)
\|_\infty = 0$$ where $\| \cdot \|_\infty$ denotes the sup-norm on $B_b(X_1)$.
\end{enumerate}

Let $E : \Sigma_1 \to \B(\H)$ be a spectral measure on $X_1$. Define a family of
maps $\{\Ah\} : \Sigma_2 \to \B(\H)$ by the formula
$$\Ah(\Delta) = \int_{X_1} p_\hbar(\Delta, \omega) \, dE(\omega)$$
for any $\Delta \in \Sigma_2$.

\begin{thm} The family $\{\Ah\} : \Sigma_2 \to \B(\H)$ defines an
ASM on $(X_2, \Sigma_2, \E_2)$. If $E$ is normalized then $\{\Ah\}$ is
also normalized.
\end{thm}

See Example \ref{ex:unshspin} below for a  concrete example of this smearing technique.

The physical interpretation (for finite systems) is that $p_\hbar$ models the
noise or uncertainty in interpreting the readings of a measurement.
For example, if $E$ has an eigenstate $\phi = E(\{\omega\}) \phi$, then the
expectation value of $\Ah(\Delta)$ when the system
is in state $\phi$ is given by
$$\< \phi | \Ah(\Delta) | \phi \>  = p_\hbar(\Delta, \omega).$$
Thus, $p_\hbar$ determines a (conditional) confidence measure of the system.

\subsection{Quasiprojectors and Semiclassical Limits }\label{ex:semi}

In this example, we show that the theory of ASMs can be used to study
semiclassical limits.
The relevant background for the
material in this section can be found in Chapters 10 and 11 of Omnes book
\cite{Omn99}.
We first need the following well-known result which is an easy consequence of
the functional calculus and spectral mapping theorem. 
(See also Lemma 5.1.6. \cite{WO93}.) It gives a rigorous statement of the
procedure used to ``straighten out'' quasiprojectors
into projections.

\begin{lem} Let  $\{a_\hbar: \hbar >0\}$ be a continuous family of
elements in a $C^*$-algebra $\B$
such that $0 \leq a_\hbar \leq 1$ for each $\hbar > 0$ and $$\lim_{\hbar \to 0}
\|a_\hbar - a_\hbar^2\| = 0.$$
There is a continuous family of projections $\hbar \mapsto p_\hbar = p_\hbar^* =
p_\hbar^2$ such that
$$\lim_{\hbar \to 0}\|a_\hbar - p_\hbar\| = 0.$$
\end{lem}

Let $(X, \Sigma, \E)$ be an asymptotic measure space.

\begin{prop} Let $\{\Ah\}$ be a normalized ASM on
$X$. For each
subset $\Delta \in \E$ there is a continuous family of projections
$\{P_\hbar(\Delta)\}$ such
that
$$\lim_{\hbar \to 0} \| \Ah(\Delta) - P_\hbar(\Delta) \| = 0.$$
Moreover if $\Delta$ and $\Delta'$ are disjoint measurable sets in $\E$ then
$$\lim_{\hbar \to 0} \| P_\hbar(\Delta) P_\hbar(\Delta') \| = 0.$$
\end{prop}

The relation to semiclassical limits occurs when we take $X$ to be the locally
compact
phase space of a classical system and $\B = \B_1(\H)$ to be the algebra of
trace-class
operators.

\begin{prop} Let $\{\Ah\}$ be a Borel ASM on $X$ with locally
compact
trace.
Then for any subset $\Delta \in \CC_X$ we have
$$\lim_{\hbar \to 0} \tr(\Ah(\Delta) - \Ah(\Delta)^2) = 0$$
and there is a unique integer $N_\Delta \in \N$ such that
$$N_\Delta = \lim_{\hbar \to 0} \tr(\Ah(\Delta)).$$
Moreover, this integer is constant on the asymptotic equivalence class of
$\{\Ah\}$.
\end{prop}

Suppose $X$ denotes the position-momentum phase space $(x, p)$ of a
particle. Let $\{\Ah\}$ be a
locally compact trace Borel ASM on $X$. A bounded rectangle $R$ in phase space
with center
$(x_0, p_0)$ and
sides $2 \Delta x$ and $2 \Delta p$ can then be used to represent a classical
property asserting
the simultaneous existence of the position and momentum $(x_0, p_0)$ of the
particle with 
given error bounds $(\Delta x, \Delta p)$ on measurement. The nonnegative
integer $N_{R}$ which
satisfies $$N_{R} = \lim_{\hbar \to 0} \tr (\Ah(R))$$
can then be interpreted as the number of semiclassical states of the particle
bound in the rectangular box $R$,
which is familiar from  elementary statistical mechanics. We then have that
$$\aligned \tr(\Ah(R) - \Ah(R)^2 ) & = N_R O(\hbar) \\
          \tr(E_\hbar(R) - \Ah(R)) & = N_R O(\hbar).
\endaligned$$ Thus, $\hbar$ represents a {\it classicity parameter}. When $\hbar
\approx 0$ is small, the 
quantum representation of the classical property is essentially correct and when
$\hbar \approx 1$ the classical property
has essentially no meaning from the standpoint of quantum mechanics. Since these
relations are preserved 
on equivalence classes, ``a classical property corresponding to a sufficiently
large a priori bounds $\Delta x$
and $\Delta p$ is represented by a set of equivalent quantum projectors''
\cite{Omn99},
i.e., equivalent locally compact trace ASMs. 
In addition, if $R_1$ and $R_2$ are disjoint rectangles, representing
distinct classical properties, then we have that
$$\| \Ah(R_1) \Ah(R_2) \| = O(\hbar)$$
and so ``two clearly distinct classical properties are (asymptotically) mutually
exclusive when considered as quantum properties'' \cite{Omn99}.

\subsection{Unsharp Spin Measurements of Spin-$\frac12$ Systems }\label{ex:unshspin}

In this example, we give a geometric classification of certain asymptotic
spectral measures associated to pure spin-$\frac12$ particles. See Example (\ref{ex:spinhalf}) for
the relevant definitions. 

Let $\H = \C^2$ be the state space of a pure spin-half system. Let $X_{\text{spin}} = \{-\frac12, +\frac12\}$ be the measurements corresponding to ``spin-down'' and ``spin-up'', respectively.

\begin{dfn}  An asymptotic spectral measure $\{\Ah\}_{\h \in (0, 1]}$
on $X_{\text{spin}}$ will be called {\it spin} if each $A_\hbar = \{\Ah^+, \Ah^-\}$ is a spin POVM as in Definition \ref{dfn:spinpovm}.
\end{dfn}

\begin{thm} There is a bijective correspondence between spin
asymptotic spectral measures $\{\Ah\}_{\h \in (o,1]}$ and 
continuous maps $\vec A : (0,1] \to B^3$ such that
$$\lim_{\hbar \to 0} \| \vec A(\hbar)\| = 1.$$
This correspondence is given by the formula
$$\Ah^\pm = \frac12(I \pm \vec A(\hbar) \cdot \vec \sigma).$$ 
\end{thm}
\smallskip

Thus, we can geometrically realize the space of spin asymptotic spectral
measures as the space of continuous
paths in the closed unit ball of $\R^3$ which asymptotically approach the  unit
sphere, i.e. they are ``asymptotically sharp.'' Note that this provides
nontrivial examples of asymptotic spectral measures which do not converge to 
a fixed spectral measure.

Let $\vec n$ be a unit vector and define $\vec A(\hbar) = (1 -
\hbar)\vec n$. The associated spin asymptotic
spectral measure given by  
$$A_h^\pm = \frac12(I \pm (1-\hbar)\vec n \cdot \vec \sigma)$$
is used by Roy and Kar \cite{RK99} to analyze eavesdropping strategies in
quantum cryptography using EPR pairs of correlated
spin-$\frac12$ particles. Violations of Bell's inequality occur when the
parameter $\hbar > 1 - \sqrt{2}(\sqrt{2}-1)^\frac12$.

This spin ASM is also obtained by the asymptotic smearing construction in Example \ref{ex:noise}.
Let
$E^\pm = A_0^\pm$ be the spectral measure
associated to the unit vector $\vec n$. Define the family $\{p_\hbar\} : \P(X_2) \times X_2
\to [0,1]$ by the formula $p_\hbar(\Delta, j) = \sum_{i \in \Delta}
\lambda^\hbar_{ij}$ where $(\lambda^\hbar_{ij})$ is the stochastic matrix
$$(\lambda^\hbar_{ij}) = \begin{pmatrix}  1 - \frac{\hbar}{2} & \frac{\hbar}{2} \\  
\frac{\hbar}{2} & 1 - \frac{\hbar}{2} \end{pmatrix}.$$
One can then verify that
$$\Ah^\pm = \sum_{l = \mp \frac12} p_\hbar(\{\pm \frac12\}, l) E^\mp.$$

\begin{cor} Spin asymptotic spectral measures $\{\Ah\}$ and
$\{B_\hbar\}$ are equivalent if and only if their 
associated maps $\vec A, \vec B : (0, 1] \to B^3$ are asymptotic, i.e., 
$$\lim_{\hbar \to 0} \|\vec A(\hbar) - \vec B(\hbar)\| = 0.$$
\end{cor}

\section{Conclusion}
In conclusion, we have reviewed the relationship between quantum theory and asymptotic morphisms in the $\cs$-algebra $E$-theory of Connes-Higson. Asymptotic measure theory, which is based on asymptotic spectral measures, bridges the gap between POVMs in quantum theory and asymptotic morphisms in $E$-theory. Examples and applications of asymptotic measure theory to various aspects of quantum theory were discussed.


\section*{Acknowledgment}
The author would like to thank his graduate student Diane Martinez for her help in preparing the original paper \cite{MT02}. The financial support for this work was provided in part by the National Science Foundation via research grant NSF DMS-0071120 and Dartmouth College.



%

\end{document}